\documentclass{ws-p9-75x6-50}
\usepackage{amssymb}

\begin{document}

\title{GRworkbench: a computational system based on differential geometry}

\author{Susan M. Scott, Benjamin J. K. Evans and Antony C. Searle}

\address{Department of Physics and Theoretical Physics, Faculty of Science,
\\
The Australian National University, Canberra ACT 0200 Australia\\E-mail:
Susan.Scott@anu.edu.au, Ben.Evans@anusf.edu.au, Antony.Searle@anu.edu.au}


\maketitle

\abstracts{We have developed a new tool for numerical work in General
Relativity: GRworkbench.  While past tools have been \emph{ad hoc},
GRworkbench closely follows the framework of Differential Geometry to provide a
robust and general way of computing on analytically defined space-times.  We
discuss the relationship between Differential Geometry and C++ classes in
GRworkbench, and demonstrate their utility.}

\section{Introduction}

We have developed a new class of computational tool for General Relativity.
Previous tools have fallen into three categories; large scale simulations
that evolve space-times from initial conditions, symbolic manipulators,
and \emph{ad hoc} numerical systems.

GRworkbench\cite{website} uses numerical variants of standard differential geometric
entities to rigorously define space-times in a fashion amenable to computation.
This system forms a strong base on which to build generally applicable
numerical algorithms, capable of acting on any space-time for which a
basic analytic definition is available.

This paper will focus on GRworkbench's roots in differential
geometry and will demonstrate the software's wide applicability to problems via
consideration of a specific example---geodesic tracing in the Schwarzschild
space-time.  For a discussion of numerical algorithms, visualization
techniques and the user interface employed by GRworkbench, see Evans\cite{bjke}
and Searle\cite{acs}.

\begin{figure}

%\figurebox{22pc}{15pc}{} % to have a box alone
%\epsfxsize=100pc % will enlarge or reduce the postscript figures based on the xsize
%\epsfbox{precessing.eps} % postscript image file name

\resizebox{\textwidth}{!}{\includegraphics{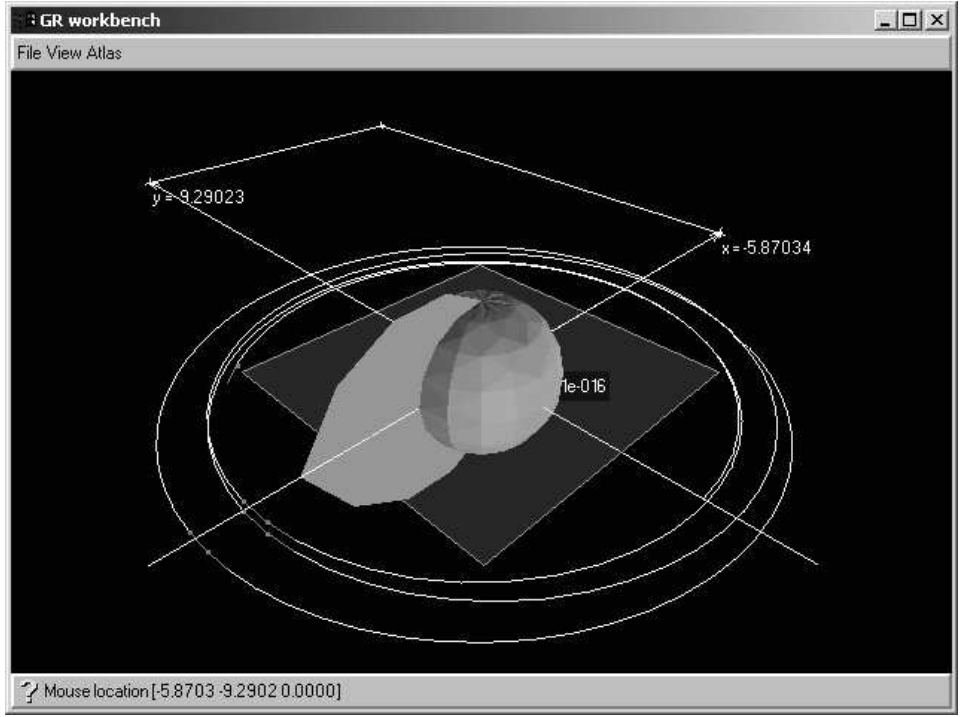}}
\caption{A precessing orbit near the event horizon of a Schwarzschild black hole ($M = 0.5$), computed and visualized in GRworkbench.  The flat
protrusion from the event horizon is the $\phi = 0, 2\pi$ boundary of the spherical polar
chart.}
\label{fig:screenshot}
\end{figure}

\section{Discrete differential geometric structure}

We follow the conventions of Hawking and Ellis\cite{hawkingandellis}:
A $C^r$ \emph{n-dimensional manifold} $\mathcal{M}$ is a set $\mathcal{M}$
together with a $C^r$ \emph{atlas} $\{(\mathcal{U}_\alpha,\phi_\alpha)\}$,
that is to say a collection of charts $(\mathcal{U}_\alpha,\phi_\alpha)$ where
the $\mathcal{U}_\alpha$ are subsets of $\mathcal{M}$ and the $\phi_\alpha$
are one-to-one maps of the corresponding $\mathcal{U}_\alpha$ to open sets in
$\mathbb{R}^n$ such that
\begin{enumerate}
\item the $\mathcal{U}_\alpha$ cover $\mathcal{M}$, i.e. $\mathcal{M} = \bigcup_\alpha\mathcal{U}_\alpha$,
\item if $\mathcal{U}_\alpha\cap\mathcal{U}_\beta$ is non-empty, then the map
\begin{displaymath}
\phi_\alpha\circ\phi_\beta^{-1}:\phi_\beta(\mathcal{U}_\alpha\cap\mathcal{U}_\beta)\rightarrow\phi_\alpha(\mathcal{U}_\alpha\cap\mathcal{U}_\beta)
\end{displaymath}
is a $C^r$ map of an open subset of $\mathbb{R}^n$ to an open subset of $\mathbb{R}^n$.
\end{enumerate}
It is assumed, as in Hawking and Ellis, that we are dealing with \emph{paracompact,
connected, $C^\infty$ Hausdorff manifolds without boundary}.

\begin{figure}
\resizebox{\textwidth}{!}{\includegraphics{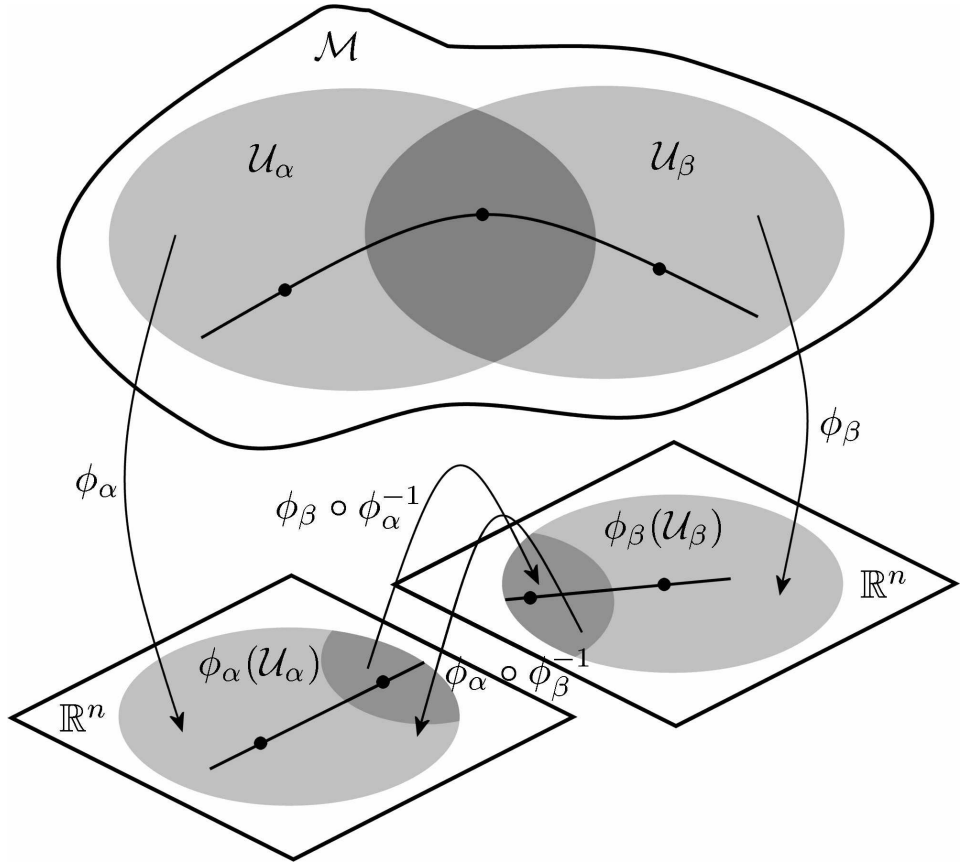}}
\caption{Schematic depiction of a geodesic crossing from the chart $(\mathcal{U}_\alpha,\phi_\alpha)$ to the chart $(\mathcal{U}_\beta,\phi_\beta)$.}
\label{fig:mappings}
\end{figure}

A computer cannot numerically represent the set $\mathcal{M}$---that is, it cannot represent
$\mathcal{M}$ using numbers.  Nor can it so represent $\mathcal{U}_\alpha$ or the mapping
$\phi_\alpha$ for any chart $(\mathcal{U}_{\alpha}, \phi_{\alpha})$.  The
manifold, subsets of the manifold, and functions on the manifold are
\emph{abstract} entities; the computer cannot deal with them \emph{numerically}.  This is not to
say that computers cannot deal with these abstract entities
\emph{symbolically}---there exist numerous symbolic manipulators, such as
Mathematica\cite{mathematica} and Sheep\cite{sheep}, for this purpose.

Computers can, however, operate numerically in $\mathbb{R}^n$.  The set
$\phi_\alpha(\mathcal{U}_\alpha)$ can be represented numerically, as can the function
$\phi_\alpha \circ \phi_\beta^{-1} : \phi_\beta(\mathcal{U}_\alpha \cap \mathcal{U}_\beta) \rightarrow \phi_\alpha(\mathcal{U}_\alpha \cap \mathcal{U}_\beta)$.
Computers
represent real numbers as \emph{floating-point} numbers.
This is analogous to base-2 scientific notation, where $(a,b)$ represents $a \times 2^b$.
On a modern computer,
real numbers are typically represented using 64 bits, meaning that the computer
can represent up to $2^{64} \approx 10^{19}$ rational numbers, spaced
approximately logarithmically
along the real line.  This means that, strictly speaking, all sets
representable by the computer
are closed, compact and totally disconnected.  Continuity cannot be sensibly
defined for functions on a totally disconnected domain, and many functions,
such as $y = \frac{2}{3}x$, are, surprisingly, \emph{not} one-to-one.  In this
example, two numbers adjacent in the representation, when multiplied by
$\frac{2}{3}$, may both produce the same floating-point result, as the
difference between the true results is less than the precision of the
discrete representation.

Thus, although a computer cannot directly represent a manifold, atlas or chart as
defined above,  it can produce \emph{similar} objects, which we
distinguish from the abstract entities by use of a \textsf{Sans Serif typeface}.

We define an \textsf{Atlas}, the numerical representation of an atlas
$\{(\mathcal{U}_{\alpha}, \phi_{\alpha})\}$, as a collection of
\textsf{Chart}s, the numerical representations of charts
$(\mathcal{U}_{\alpha}, \phi_{\alpha})$.  We define the \textsf{Chart}
representing $(\mathcal{U}_{\alpha}, \phi_{\alpha})$ as:
\begin{enumerate}
\item The numerical representation of the set $\phi_\alpha(\mathcal{U}_\alpha)$.
\item The set of numerical representations of functions mapping from the
\textsf{Chart} to other \textsf{Chart}s:
$\{\phi_\beta \circ \phi_\alpha^{-1} : \phi_\alpha(\mathcal{U}_\alpha \cap \mathcal{U}_\beta) \rightarrow \phi_\beta(\mathcal{U}_\alpha \cap \mathcal{U}_\beta)\}$.
\end{enumerate}
We cannot define a \textsf{Manifold} as an \textsf{Atlas} combined with the
set of equivalence classes of
points $x \in \phi_\alpha(\mathcal{U}_\alpha)$ under the mappings
$\phi_\beta \circ \phi_\alpha^{-1}$, as these mappings are not, in general,
one-to-one, due to the discrete representation employed by the computer.  It is possible that
$\phi_\alpha \circ \phi_\beta^{-1}(\phi_\beta \circ \phi_\alpha^{-1}(x)) \not= x$,
though the difference should be ``small'', that is, on the order of the local
resolution $\epsilon$ of the discrete representation.  In general,
there appears to be no sensible definition for a \textsf{Manifold}, and we do not adopt one.

Although we cannot construct an authoritative set $\mathcal{M}$ using the above
procedure, we
can still produce a \emph{useful} definition of a \textsf{Point} as the numerical
representation of $p \in \mathcal{M}$.  First define a \textsf{Coordinate}\footnote{Internally named
a \textsf{Node} by GRworkbench for historical reasons.} by:
\begin{enumerate}
\item The numerical representation (\textsf{Chart}) of a chart $\alpha$ (abbreviated notation for $(\mathcal{U}_\alpha,\phi_\alpha)$).
\item The numerical representation of a point $x \in \phi_\alpha(\mathcal{U}_\alpha)$.
\end{enumerate}
Note that the numerical representation of $x$ is simply an $n$-tuple of real numbers,
so it is necessary to additionally specify the chart in order to give those numbers the
context of a mapping.
We can identify a \textsf{Coordinate} $(\alpha, x)$ with a point of the
manifold, $\phi_\alpha^{-1}(x) \in \mathcal{U}_\alpha \subseteq \mathcal{M}$.
We do not, however, have a numerical representation of the function
$\phi_\alpha^{-1}$, so to define a \textsf{Point} representing $p \in  \mathcal{U}_\alpha$
we use a set of \textsf{Coordinate}s:
\begin{enumerate}
\item The \textsf{Coordinate} $(\alpha, \phi_\alpha(p) = x)$. 
\item The \textsf{Coordinate}s $\{(\beta, \phi_\beta \circ \phi_\alpha^{-1}(\phi_\alpha(p))) : \beta \not= \alpha\}$.
\end{enumerate}
Note that these are \emph{not} the \textsf{Coordinate}s $(\beta, \phi_\beta(p))$,
though the difference should be ``small''.

Thus, although we can define a \textsf{Point}, its definition is tied to its
coordinates under a particular chart.  If we were to produce
\textsf{Coordinate}s $(\alpha, \phi_\alpha(p))$ and
$(\beta, \phi_\beta(p))$ for the same point
$p \in \mathcal{U}_\alpha \cap \mathcal{U}_\beta \subseteq \mathcal{M}$, the
\textsf{Point}s manufactured from them would not, in general, consist of precisely the
same \textsf{Coordinate}s (condition 2 above), and thus would \emph{not} be equal so far as the
computer is concerned.  We cannot produce a chart-independent representation
of a point; we can only produce a chart-dependent \textsf{Point} that is
``approximately'' chart-independent, in the sense that the difference between
the representations arising from different charts is ``small''.

There is an obvious similarity between the \textsf{Atlas} and \textsf{Point};
both are collections of \emph{different} numerical representations of the \emph{same}
abstract object.  This abstract layer identifying different numerical
representations imparts (approximate) \emph{chart-independence} to numerical operations
performed within the structure.  Essentially, in the midst of performing
a computation the computer can change charts as required by mapping the pertinent
parameters to another \textsf{Chart}.

An example we will follow throughout this paper is the computation of a discrete
approximation to a geodesic.  A geodesic $\gamma : \mathbb{R} \rightarrow \mathcal{M}$
may, for some interval around $\tau \in \mathbb{R}$, lie in the domain $\mathcal{U}_\alpha$ of
a chart $\alpha$.  It can be represented numerically by the series of \textsf{Coordinate}s
(and thus by the series of \textsf{Point}s generated from the \textsf{Coordinate}s)
$(\alpha, \phi_\alpha(\gamma(\tau))), (\alpha, \phi_\alpha(\gamma(\tau + \delta))), (\alpha, \phi_\alpha(\gamma(\tau + 2\delta))), \dots$
for some \emph{step-size} $\delta > 0$, but it is possible that for some $k \in \mathbb{N}$,
$\gamma(\tau+(k + 1)\delta) \not\in \mathcal{U}_\alpha$.  In this case, for $\delta$ chosen to be sufficiently small, there
exists some \textsf{Chart} $\beta$ such that $\gamma(\tau+(k + 1)\delta) \in \mathcal{U}_\beta$, and
$\gamma(\tau+k\delta) \in \mathcal{U}_\alpha\cap\mathcal{U}_\beta$.  Without
the abstract identification of the \textsf{Coordinate} $(\alpha, \phi_\alpha(\gamma(\tau+k\delta)))$
with the \textsf{Coordinate} $(\beta, \phi_\beta \circ \phi_\alpha^{-1}(\phi_\alpha(\gamma(\tau+k\delta))))$
via the \textsf{Point} representing the point $\gamma(\tau+k\delta)$,
there would be no way in which the next \textsf{Point}, representing the point $\gamma(\tau+(k + 1)\delta)$, could be computed.  With
the identification, however, we can continue to compute the numerical representation
from algorithms operating on \textsf{Chart} $\beta$.  This procedure is depicted schematically in Figure \ref{fig:mappings}.

The interest of users in entities, such as geodesics, that are structured sets
of \textsf{Point}s, motivates us to define the \textsf{Object}.  An \textsf{Object}
is a set of \textsf{Point}s.  The \textsf{Point}s constituting an \textsf{Object}
will have \textsf{Coordinate}s.  We define a \textsf{Segment} for a certain
\textsf{Object} and \textsf{Chart} as the \emph{image} of the set represented
by the \textsf{Object} on that \textsf{Chart}.  As such, a \textsf{Segment}
is a set of \textsf{Coordinate}s.

To use the differential geometric framework to represent a space-time, we must
provide one additional component.  We extend the definition of a \textsf{Chart}
to encompass the provision of a \emph{metric} $\mathbf{g}$---a $(0,2)$ symmetric tensor field
on $\mathcal{M}$.  We require that each \textsf{Chart} provide a function returning the
tensor components $g_{ij}|_x$ with respect to the chart's coordinate basis
$\{\partial/\partial x^a\}$, for any coordinate $x \in \phi_\alpha(\mathcal{U}_\alpha)$.

\section{Implementation}

\begin{figure}
\resizebox{\textwidth}{!}{\includegraphics{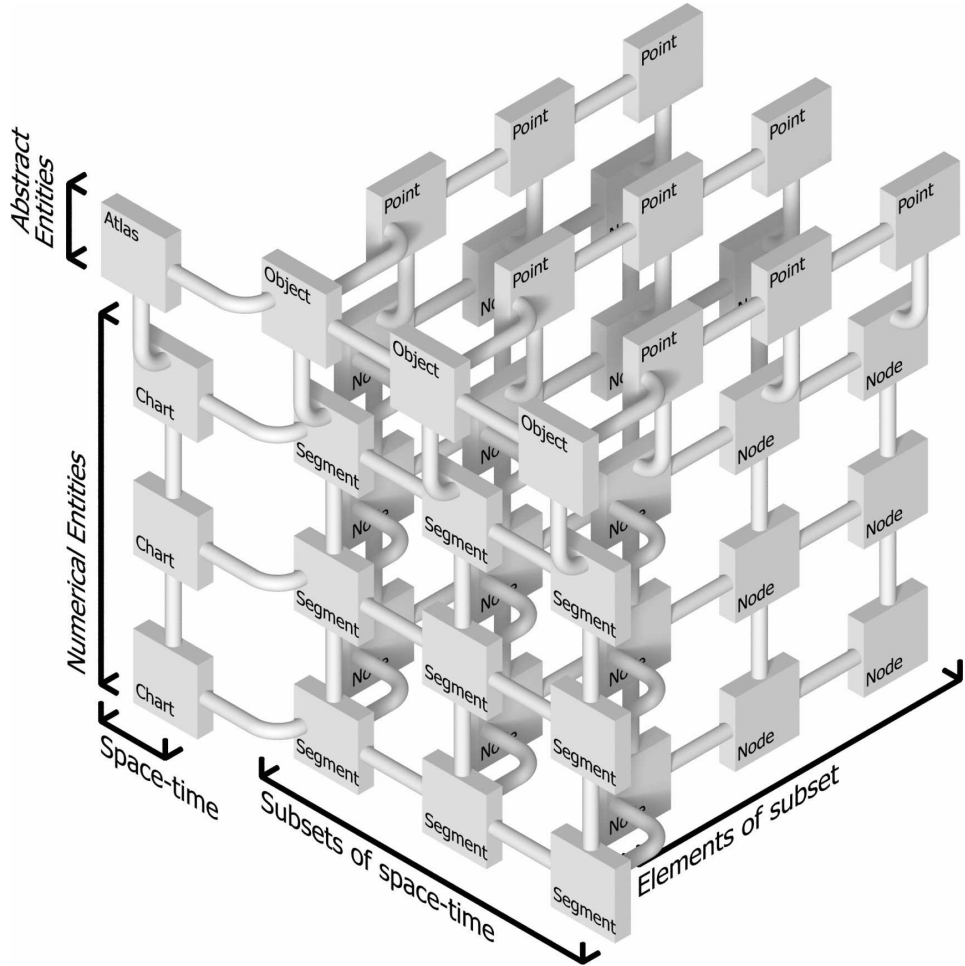}}
\caption{Differential geometric structure of a manifold as encoded by GRworkbench.  The
entities labelled \textsf{Node}s are generalisations of the
\textsf{Coordinate} concept that may additionally carry a vector
and/or scalar.}
\label{fig:3d}
\end{figure}

The numerical differential geometric system defined above may be naturally
expressed as a collection of \emph{classes} in the  C++\cite{stroustrup}
programming language.  A class is itself a collection of data and related functions.
The notation \textsf{A::B} indicates that \textsf{B} is a \emph{member} of
class \textsf{A}.

The \textsf{Atlas} class stores chart-independent data and a list of 
the \textsf{Charts} $\{(\mathcal{U}_{\alpha}, \phi_{\alpha})\}$
that comprise it.  In the case of the \textsf{Atlas}
representing the Schwarzschild space-time, the \textsf{Atlas} stores
the Schwarzschild mass $M$ that defines certain properties of the
\textsf{Chart}s, such as mandating that the radius $x_1 > 2M$ for exterior
charts.

The \textsf{Chart} representing $(\mathcal{U}_{\alpha}, \phi_{\alpha})$ must provide
\begin{enumerate}
\item The set $\phi_{\alpha}(\mathcal{U}_\alpha) \subseteq \mathbb{R}^n$.
\item The functions $\{\phi_\beta \circ \phi_\alpha^{-1} : \phi_\alpha(\mathcal{U}_\alpha \cap \mathcal{U}_\beta) \rightarrow \phi_\beta(\mathcal{U}_\alpha \cap \mathcal{U}_\beta)\}$.
\item The metric tensor $g_{ij}|_x$, for $x \in \phi_\alpha(\mathcal{U}_\alpha)$.
\end{enumerate}
For maximum flexibility, we require the user to define the set $\phi_\alpha(\mathcal{U}_\alpha)$
in terms of a \emph{Boolean} function on $\mathbb{R}^n$, that is, a function
\textsf{Chart::Interior} $: \mathbb{R}^n \rightarrow \{\mbox{true}, \mbox{false}\}$.
\begin{equation}
\mbox{\textsf{Chart::Interior}}(x) = \left\{\begin{array}{ll}
\mbox{true}  & \mbox{if $x \in \phi_\alpha(\mathcal{U}_\alpha)$} \\
\mbox{false} & \mbox{otherwise}
\end{array}\right.
\end{equation}
This formalism allows the user to plug in any algorithm to define the
domain of the \textsf{Chart}.

For an exterior spherical polar chart of the Schwarzschild space-time, the
\textsf{Chart::Interior} function is defined as
\begin{equation}
\mbox{\textsf{Chart::Interior}}(x) = \left\{\begin{array}{ll}
\mbox{true}  & \mbox{if $x_1 > 2M$ and $0 < x_2 < \pi$ and $0 < x_3 < 2\pi$} \\
\mbox{false} & \mbox{otherwise}
\end{array}\right.
\end{equation}

Maps to other charts, $\phi_\beta \circ \phi_\alpha^{-1}$, are implemented as
$\mathbb{R}^n \rightarrow \mathbb{R}^n$ functions.  An obvious disadvantage of
the need to define maps $\phi_\beta \circ \phi_\alpha^{-1}$ rather than simply maps
$\phi_\alpha$ is that, in general, an \textsf{Atlas} comprising $m$ \textsf{Charts}
requires up to $m(m-1)$ $\mathbb{R}^n \rightarrow \mathbb{R}^n$ maps rather than just the
$m$ $\mathcal{M} \rightarrow \mathbb{R}^n$ maps.  This number can be reduced
by permitting the implicit definition of maps, by application of several
maps between other charts, such as the map
$\phi_\beta \circ \phi_\alpha^{-1} = (\phi_\beta \circ \phi_\gamma^{-1}) \circ
(\phi_\gamma \circ \phi_\alpha^{-1}) :
\phi_\alpha(\mathcal{U}_\alpha \cap \mathcal{U}_\beta \cap \mathcal{U}_\gamma)
\rightarrow
\phi_\beta(\mathcal{U}_\alpha \cap \mathcal{U}_\beta \cap \mathcal{U}_\gamma)$.

Two spherical polar charts are required to cover the entire exterior ($x_1 > 2M$) region of the
Schwarzschild space-time.  Choosing the second spherical polar chart so that its polar
axis corresponds to the coordinates\footnote{
We use the C++ convention that indices begin from $0$.
} $x_2=\pi/2$ and $x_3=\pi/2,3\pi/2$ allows us to define the map between the charts as: 
\begin{equation}
\begin{array}{ll}
x'_0 & = x_0 \\
x'_1 & = x_1 \\
x'_2 & = \arg(\sin x_2\sin x_3 + \sqrt{\sin^2x_2\sin^2x_3 - 1}) \\
x'_3 & = \arg(-\sin x_2\cos x_3 + \sqrt{-1}\cos x_2)
\end{array}
\end{equation}
This map is its own inverse.

The numerical representation of a differential geometric structure cannot
know, in advance, if that representation is complete.  That is, the implementation cannot
determine if the \textsf{Atlas} provided covers the entire manifold $\mathcal{M}$, as the
implementation has no \emph{a priori} knowledge of the manifold $\mathcal{M}$ itself.  Neither can the
implementation determine if the $\mathbb{R}^n \rightarrow \mathbb{R}^n$
maps supplied by the \textsf{Chart}s are comprehensive, as the relationships
between \textsf{Chart}s are defined solely in terms of these maps.  This allows the user freedom
to implement systems for which the \textsf{Atlas} provided does \emph{not}
cover the manifold, and for which the $\mathbb{R}^n \rightarrow \mathbb{R}^n$
maps supplied by the \textsf{Chart}s are \emph{not} comprehensive.

This is not cause for concern.  When the implementation cannot successfully
continue correct computation due to the lack of a \textsf{Chart} or map, the
algorithm halts.  If the user wishes to perform the computation, they
must add a new component to the incomplete system.  In many circumstances,
though, the user may be concerned only with a portion of the system.  For
example, to study orbits in a Schwarzschild space-time, it is sufficient
to define only \emph{exterior} charts.  To require the user to define portions
of the space-time which they have no intention of utilizing---in this case, the interior Schwarzschild space-time---would be to
unnecessarily inconvenience the user.

To complete the definition of the \textsf{Chart}, the user must supply a
function returning the metric tensor components $g_{ij}|_x$, for $x \in \phi_\alpha(\mathcal{U}_\alpha)$.  For any exterior
spherical polar chart of the
Schwarzschild space-time, the components are defined as:
\begin{equation}
\begin{array}{llll}
g_{00}|_x = -1 + \frac{2M}{x_1}\ & 
g_{01}|_x = 0 & 
g_{02}|_x = 0 & 
g_{03}|_x = 0 \\
g_{10}|_x = 0 & 
g_{11}|_x = (1 - \frac{2M}{x_1})^{-1}\ & 
g_{12}|_x = 0 & 
g_{13}|_x = 0 \\
g_{20}|_x = 0 & 
g_{21}|_x = 0 & 
g_{22}|_x = (x_1)^2\ & 
g_{23}|_x = 0 \\
g_{30}|_x = 0 & 
g_{31}|_x = 0 & 
g_{32}|_x = 0 & 
g_{33}|_x = (x_1)^2\sin^2x_2\ 
\end{array}
\end{equation}

In the current implementation, \textsf{Atlas}es and \textsf{Chart}s are defined by the user
by modifying trivial C++ code in supplied ``template'' files\footnote{Not to
be confused with the C++ \textsf{template} keyword.}.  Only the most
basic programming skills are required of the user.  Compared to an obvious
alternative, namely, using a scripting language, this method has the advantage of producing
efficient, pre-compiled code, but the disadvantage of \emph{relinking} the
\emph{executable} when space-time definitions are modified.
Such modifications are, however, comparatively rare for common usage.

The remaining classes, \textsf{Object}s, \textsf{Segment}s, \textsf{Point}s
and \textsf{Coordinate}s are implemented in a straightforward way.
\textsf{Coordinate}s contain an $n$-tuple of floating point values and an
indirect reference (via their containing \textsf{Segment}) to their
\textsf{Chart}s.
Any \textsf{Point} owns a list of its \textsf{Coordinate} representations on
all applicable \textsf{Chart}s.  A \textsf{Segment} similarly maintains a
list of the \textsf{Coordinate}s of the \textsf{Point}s of a particular \textsf{Object} on a
particular \textsf{Chart}.  An \textsf{Object} maintains a list of its \textsf{Point}s, and
\textsf{Segment}s on various \textsf{Chart}s, and a \textsf{Chart} maintains a list
of \textsf{Segment}s on it from various \textsf{Object}s.  \textsf{Segment}s
and \textsf{Coordinate}s thus belong to two lists, which are represented
perpendicularly to one another in Figure \ref{fig:3d}.

The above components are genuinely sufficient to define a space-time.  Where
GRworkbench requires certain information to perform a task, such as the
geodesic equation to compute a geodesic, it extracts that information numerically
using  the relevant definition.  For example, the geodesic equation is given by
\begin{equation}
\frac{\mathrm{d}^2x^a}{\mathrm{d}\tau^2} =
-\Gamma^a_{\ bc}\frac{\mathrm{d}x^b}{\mathrm{d}\tau}
\frac{\mathrm{d}x^c}{\mathrm{d}\tau},
\end{equation}
where the Christoffel symbol $\Gamma$ is given by
\begin{equation}
\Gamma^a_{\ bc} = \frac{1}{2}g^{ad}
\left(
\frac{\partial}{\partial x^c}g_{db} +
\frac{\partial}{\partial x^b}g_{dc} -
\frac{\partial}{\partial x^d}g_{bc}
\right).
\end{equation}
We may compute $g^{ab}$ as the inverse matrix of $g_{ab}$, and compute the partial
derivatives $\frac{\partial}{\partial x^c}g_{ab}$ by numerical differentiation
of $g_{ab}$.  Thus, GRworkbench can compute the geodesic equation directly from
the user-supplied metric for any given space-time.

\section{Utility}

\begin{figure}
\resizebox{\textwidth}{!}{\includegraphics{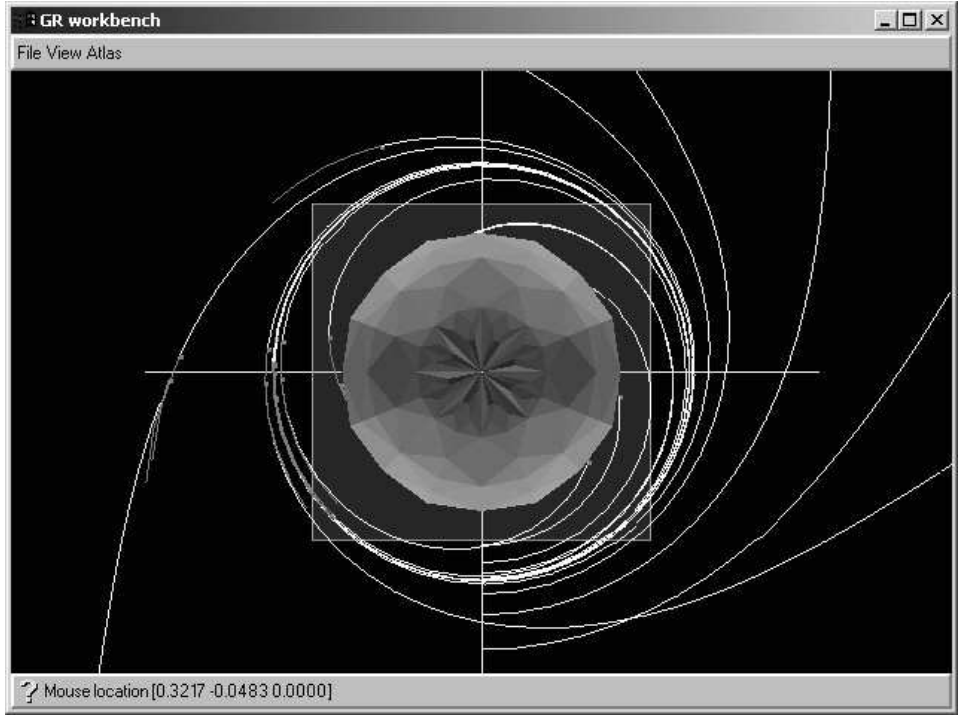}}
\caption{Null geodesics in the Schwarzschild space-time, iterating in on the
$r=1.5=3M$ circular null orbit.}
\label{fig:orbits}
\end{figure}

Using the definition for a Schwarzschild \textsf{Atlas} given in Section 3, consisting of
two exterior spherical polar \textsf{Chart}s, we can begin to trace
geodesics, such as those of Figure \ref{fig:screenshot}, on the space-time
\emph{without deriving the geodesic equations}.

A well-known fact about the Schwarzschild space-time is that there exists a
circular null orbit at $x_1=3M$.  Although this fact is not immediately apparent
analytically, it can be reproduced using GRworkbench in an \emph{experimental}
mode.

We release geodesics from $x_0 = 0, x_2 = \pi/2, x_3 = \pi/2$ for varying
values of $x_1 > 2M$, and mandate that
$\frac{\partial x^0}{\partial\tau} > 0,
\frac{\partial x^1}{\partial\tau} = \frac{\partial x^2}{\partial\tau} = 0,
\frac{\partial x^3}{\partial\tau} > 0$ and the geodesic be null.  GRworkbench
then uses the metric to produce an initial null tangent vector satisfying these
conditions\footnote{GRworkbench does this by breaking the given vector into
a purely time-like and purely space-like component, and re-scaling these parts
so that they sum to a null vector.}.  For any $x_1$, the geodesic will
either escape to infinity, or fall into the event horizon.  Using these
conditions, we can iterate down on the value of $x_1$ that divides these
two types of behaviour.  Figure \ref{fig:orbits} shows the geodesics traced in this
process.  As expected, the value below which geodesics fall into the event horizon,
and above which geodesics escape to infinity, is given by $x_1 = 1.5 = 3 \times 0.5 = 3M$
to high accuracy.  The computations were performed interactively in a few
minutes on an inexpensive notebook computer.

While this is a trivial example, recall that this functionality is available
purely from a minimal definition of the space-time---and as such, is available
for \emph{any} space-time that can be so defined.  Additionally,
there are many algorithms implemented, beyond the featured example of geodesic tracing, that
operate with similar flexibility; from the production of null cones to the
examination of causal connections, to the computation of proper distances
between points on a space-like hypersurface.  GRworkbench's firm basis
in differential geometry makes it a truly general tool, and encourages a new
\emph{experimental} approach to problems in General Relativity---one of trial and
observation.

\section{Conclusion}

GRworkbench successfully implements a numerical analogue of a standard
differential geometric system, mirroring a manifold and atlas of charts
with the C++ classes \textsf{Atlas} and \textsf{Chart}.  Provision of
a metric on a \textsf{Chart} completes the definition of a space-time
in General Relativity.

From this minimal definition, complex algorithms, such as geodesic tracing,
may be ``bootstrapped'' through layers of numerical operations, to
produce useful results, in novel ways, with minimal effort.  As such,
GRworkbench permits, and encourages, an ``experimental'' approach to
problem-solving in General Relativity.

\end{document}